\documentclass[fleqn,usenatbib]{mnras}

\usepackage{newtxtext,newtxmath}
\usepackage{breqn}

\usepackage[T1]{fontenc}
\usepackage{ae,aecompl}

\usepackage{graphicx}	
\usepackage{amsmath}	
\usepackage{xcolor}
\usepackage{float}

\title[Evidence of dayside clouds?]{Another look at the dayside spectra of WASP-43b and HD 209458b: are there scattering clouds?}

\author[Taylor $\&$ Parmentier]{
Jake Taylor$^{1,2}$\thanks{E-mail: jake.taylor@physics.ox.ac.uk} $\&$
Vivien Parmentier$^{1,3}$
\\
$^{1}$Department of Physics, University of Oxford, Parks Rd, Oxford, OX1 3PU, UK \\
$^{2}$Institut Trottier de recherche sur les exoplanètes and Département de Physique, Université de Montréal, 1375 Avenue Thérèse-Lavoie-Roux,\\
Montréal, QC, H2V 0B3, Canada\\
$^{3}$Université Côte d’Azur, Observatoire de la Côte d’Azur, CNRS, Laboratoire Lagrange, France \\
}

\date{Accepted 2023 July 15. Received 2023 July 15; in original form 2022 March 02.}

\pubyear{2021}

\begin{document}
\label{firstpage}
\pagerange{\pageref{firstpage}--\pageref{lastpage}}
\maketitle

\begin{abstract}
The search for clouds on the dayside of hot Jupiters has been disadvantaged due to hot Jupiters having a limited number of high quality space-based observations. To date, retrieval studies have found no evidence for grey clouds on the dayside, however none of these studies explored the impact of scattering clouds. In this study we reanalyse the dayside emission spectrum of the hot Jupiter WASP-43b considering the different Spitzer data in the literature. We find that, in 2 of the 4 data sets explored, retrieving with a model that contains a scattering cloud is favoured over a cloud free model by a confidence of 3.13 - 3.36 $\sigma$. The other 2 data sets finds no evidence for scattering clouds. We find that the retrieved H$_2$O abundance is consistent regardless of the Spitzer data used and is consistent with literature values. We perform the same analysis for the hot Jupiter HD 209458b and find no evidence for dayside clouds, consistent with previous studies.

\end{abstract}




\section{Introduction} \label{sec:intro}


Studying the eclipse spectra of exoplanets lets us understand the dayside environment of the planet, such as the thermal structure, chemical composition and cloud structure. The first uniform study of the dayside of exoplanets was performed in \citet{Line2014}, however the 7 out of the 9 targets that were analysed only had photometric observations. Their study neglects the inclusion of clouds citing the reasons presented in \citet{Madhu2009}, these include: clouds forming too deep in the atmosphere to impact the emergent spectra \citep{Fortney2006ApJ} and fast sedimentation rates in radiative atmospheres \citep{Barman2005}.


There has been amounting evidence in the literature from the study of the transmission spectra of exoplanets that they are indeed cloudy objects. \citet{Sing2016} perform a comparative study of 10 hot Jupiters and find evidence for clouds/aerosols due to the muted water features observed. Two of these planets were studied in great detail by \citet{Barstow2020}, they found that the transmission spectra of HD189733b and HD209458b both contain clouds and that the retrieved water abundance is robust against the cloud modelling choice used. 

The confirmation of clouds in the transmission spectra of hot Jupiters poses the question: are there clouds on their dayside? Theoretically clouds have been predicted to form on the dayside of hot Jupiters \citep{Parmentier2021,Roman2021}, clouds can inform us about the atmospheric circulation, with cloudless atmospheres having efficient day-to-night heat transport. In this study we reanalyse the eclipse spectra of the two hot Jupiters: HD 209458b \citep{Charbonneau2000} and WASP-43b \citep{Hellier2011}. There have been multiple studies of HD 209458b which require clouds to explain the transmission spectrum \citep{MacDonald2017,Pinhas2019,Barstow2020}, therefore we expect clouds to be on the terminators, could they be on the dayside too? The same cannot be said for WASP-43b, as previous work interpreting the transmission spectrum concludes that a cloud free model is sufficient \citep{Kreidberg2014}.

The eclipse spectra of both of these planets have previously been studied, however the cloud models used were only simple grey clouds, they did not model any impact scattering from the clouds could have \citep{Kreidberg2014,Lowe2014,Line2016}, we therefore reanalyse the dayside spectra of these planets taking into account that clouds can scatter light in the infrared to see if has an impact \citep{Taylor2020b}.

The structure of this manuscript is as follows: In Section 2 we outline our modelling framework and data used, Section 3 the results of our retrieval analysis and Section 4 we make our conclusions.

\section{Method} \label{sec:method}
\subsection{Model description}
We use the \textbf{N}on-linear optimal \textbf{E}stimator for \textbf{M}ultivariat\textbf{E} spectral analy\textbf{SIS} code \textsc{NEMESIS} \citep{irwin2008nemesis} to compute our model spectra. \textsc{NEMESIS} uses the correlated-$k$ approach \citep{lacis1991description} to model the spectra, which has been shown to be effective and accurate when compared to using the line-by-line or cross-section approaches \citep{garland2019effectively}. For this study the molecules used are: H$_2$O \citep{polyansky2018exomol}, CO \citep{li2015rovibrational}, CH$_4$ \citep{yurchenko2014exomol}, CO$_2$ \citep{Yurchenko2020} and NH$_3$ \citep{Coles2019} which were formatted using the techniques presented in \citet{Chubb2020}.  Our choice of molecules are used to be consistent with the modelling work of \citet{Feng2020}, as they perform a comprehensive study of the WASP-43b phase curve, however without considering scattering clouds. As we are modelling an atmosphere which is H$_2$-dominated we also modelled the H$_2$-H$_2$ and H$_2$-He collisionally-induced absorptions \citep{richard2012new}. \textsc{NEMESIS} calculates multiple scattering using the matrix operator method \citep{plass1973matrix}. The scattering component is described in more detail in \citet{Taylor2020b}. We use a parameterised temperature-pressure profile developed by \citet{Parmentier2014} which balances incoming shortwave radiation with the outgoing longwave radiation, details of its implementation into \textsc{NEMESIS} can be found in \citet{Taylor2020a} and \citet{Taylor2020b}.

\subsection{Data}
For both exoplanets in this study we have used data from Hubble/WFC3 and the Spitzer Space Telescope. 

\subsubsection{WASP-43b}
The WFC3 observations are taken from \citet{Kreidberg2014} $\&$ \citet{Stevenson2014}, and the Spitzer data are taken from \citet{Stevenson2017} thus being consistent with data used in other atmospheric retrieval studies of this exoplanet \citep{Feng2020,Irwin2020,Changeat2021}. However, there has been some discrepancy in the reduction of the Spitzer data points, therefore we also perform analysis with the Spitzer data from \citet{Blecic2014}, \citet{Morello2019} and \citet{May2020}. Hence, we perform four different analyses for WASP-43b, each use the same Hubble/WFC3 data however varying the Spitzer data. The analyses are described as followed for clarity:
\begin{enumerate}
    \item \textit{Stevenson}, where we use the Spitzer data from \citet{Stevenson2017}.
    \item \textit{Blecic}, where we use the Spitzer data from \citet{Blecic2014}.
    \item \textit{Morello}, where we use the Spitzer data from \citet{Morello2019}. 
    \item \textit{May}, where we use the 4.5$\mu$m data point from \citet{May2020} and the 3.6$\mu$m data point from \citet{Stevenson2017}. This is due to \citet{May2020} not reanalysing the 3.6$\mu$m observations with their technique as they found time variability in the observations.
\end{enumerate}

\begin{figure}
    \centering
    \includegraphics[width=0.45\textwidth]{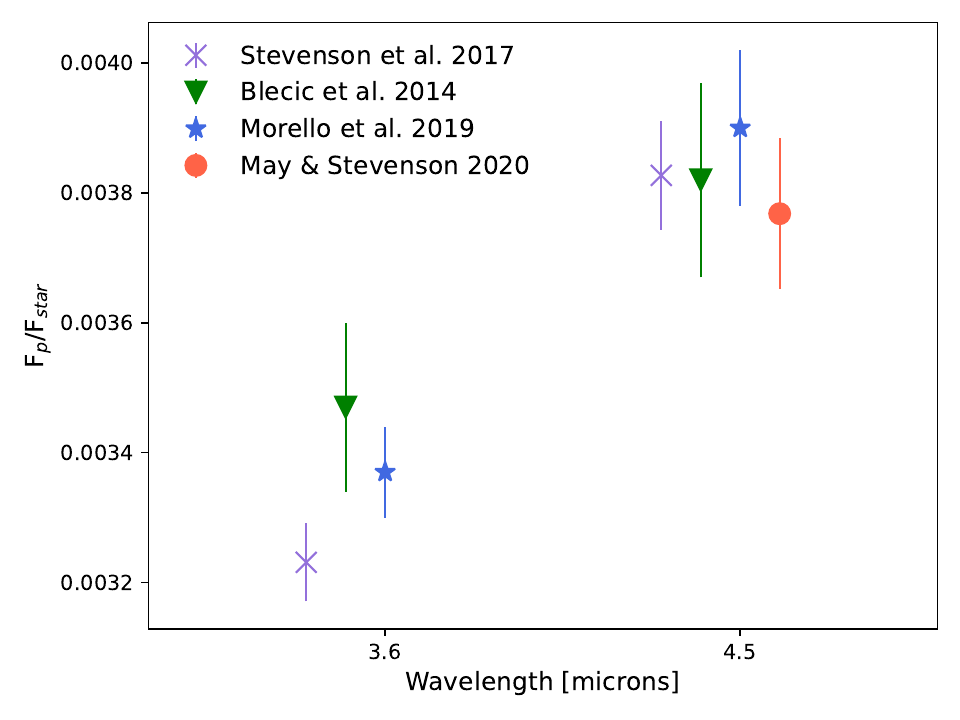}
    \caption{The 3.6 and 4.5 micron Spitzer IRAC measurements used in this study. The purple cross is the data from \citet{Stevenson2017}, the upside down green triangle is the data from \citet{Blecic2014}, the blue star is the data from \citet{Morello2019} and in red circle is the data from \citet{May2020}.}
    \label{fig:wasp-43b_data}
\end{figure}

We show in Figure \ref{fig:wasp-43b_data} the WASP-43b Spitzer data used so that the difference in eclipse depths from the different Spitzer data reductions are explicitly shown. It can be seen that there are significant differences between the eclipse depths obtained from the 3.6 micron observations.

\subsubsection{HD 209458b}
The WFC3 observations are taken from the analysis conducted in \citet{Line2016}, the Spitzer data is obtained from \citet{Lowe2014}. We split the analysis into two parts:
\begin{enumerate}
    \item We perform the analysis with 3.6 and 4.5 micron photometric points
    \item We perform the analysis with 3.6, 4.5, 5.6 and 7.8 micron points.
\end{enumerate}

\subsection{Retrieval description}
For the retrievals conducted in this study we have wrapped the forward-modelling component of \textsc{NEMESIS} in a Bayesian framework, namely using a nested sampling approach \citep{feroz2008multimodal,feroz2013importance}, which we implemented using \textsc{pymultinest} \citep{buchner2014x}. We use a nested sampling approach due to its ability to produce posterior distributions of the parameters and also for its efficiency in calculating the Bayesian evidence, which can be used in model comparison and selection.

We explore 3 different atmospheric setups. Each fit for the same molecules and parameterise the thermal structure in the same way, they only vary in their cloud setup described as follows:
\begin{enumerate}
    \item A cloud free atmosphere shown in blue in our plots
    \item An atmosphere which has a grey cloud and no scattering component. It is parameterised by fitting for the opacity of the cloud and fractional scale height. The cloud base is fixed to be at the bottom of the atmosphere. By ``grey'' we refer to the fact the opacity of the cloud is constant with wavelength and only absorbing.
    \item An atmosphere which has a scattering cloud as described in \citet{Taylor2020b}. Like the grey cloud, we fit for the opacity of the cloud, the fractional scale height and fix the cloud base to be at the bottom of the atmosphere. We also fit for an additional parameter, the single scattering albedo. We originally used the 3-point method \footnote{Fitting for 2 single scattering albedo points and a wavelength where these values transition.} as this technique has been shown to approximate the shape that various cloud condensing species single scattering albedo spectra can make, however there was no evidence to use this more complex modelling setup. The scattering is isotropic in nature.
\end{enumerate}

We present the parameters used and our prior ranges in Table \ref{tab:priors}.
\begin{table*}
    \centering
    \begin{tabular}{l|l|l|l|l|l}
    \hline
     Molecules & Priors & Temperature & Priors & Cloud & Priors\\
     \hline
     log(H$_2$O)    &    -12 to -1  & log$\kappa_{\text{IR}}$ & -4 to 1 & Single Scattering Albedo            & 0 to 1  \\
     log(CO)        &    -12 to -1  & log$\gamma_{\text{v1}}$ & -4 to 3 & log(Opacity)      & -2 to 7\\
     log(CO$_2$)    &    -12 to -1  & Irradiated Temperature           & 400 to 3000 & Fractional Scale Height (FSH)            & 0 to 1\\
     log(CH$_4$)    &    -12 to -1  \\
     log(NH$_3$)    &    -12 to -1  \\
     \hline
    \end{tabular}
    \caption{Molecules, temperature parameters and cloud parameters used in the free retrieval analysis alongside the assumed prior ranges for which the Bayesian algorithms can explore.}
    \label{tab:priors}
\end{table*}
We then consider the data in two different ways, the first is without any offset between WFC3 and Spitzer and the second is by fitting for an offset. We fit for an offset as the instruments are not necessarily calibrated to be consistent with each other. It has been shown that fitting for an offset helps to resolve this issue, however ideally overlapping spectral regions is the ideal scenario \citep{Yip2021}. To assess the impact of the Spitzer points, we add a shift parameter to the WFC3 observations. We find that in no case is there evidence for needing an offset parameter, therefore do not present these analyses in this manuscript. We present the parameters used and our prior ranges in Table \ref{tab:priors}. We use the system parameters for WASP-43b and HD 209458b from \citet{Stevenson2017} and \citet{Line2016} respectively.

\subsection{Bayesian Evidence and Model Selection}

We use Bayesian model comparison to evaluate the statistical significance of one retrieval model over another \citep[e.g.][]{Ford2007,schulze2012bayesian}.
We compute the Bayes factor $\ln B = \ln Z_2 - \ln Z_1$, where $Z_1$ and $Z_2$ are the evidences for the cloud free and scattering cloud models, respectively, as calculated by the nested sampling algorithm. If the Bayes factor is significantly less than unity ($\ln B<0$), the data supports model 1 (cloud free model). If it is significantly greater than unity, the reverse is true. If $|\ln B|$ is close to zero, one cannot discriminate between the two models. It is common practice, when using the Bayes factor for model comparison, to interpret the results using the adapted Jeffrey scale \citep{kass1995bayes}. For example, the preference for model 2 over model 1 is considered statistically significant (``substantial'') only if $\ln B > 3.2$. We use the formulation in \citet{Trotta2008} to convert the Bayes factor into a sigma confidence.

\section{Results} \label{sec:results}
In this section we will describe the results of the retrieval analysis and split the results by each planet. First HD 209458b and then WASP-43b, as this target required a more detailed analysis.

\subsection{HD 209458b}
\begin{figure*}
    \centering
    \includegraphics[width=\textwidth]{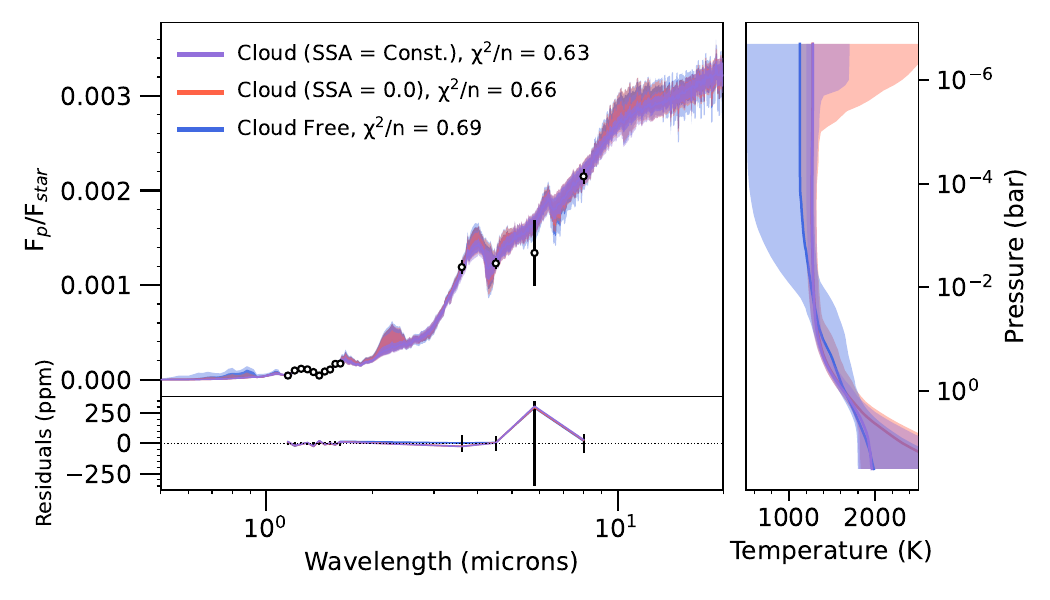}
    \caption{Left: Best fitting models of HD 209458b. We present the Scat. Cloud, Grey Cloud and Cloud Free models in purple, red and blue respectively. We present the reduced $\chi^2$ for each model inset. The bottom panel shows the residual of the fit, with the grey vertical lines representing the error on the measurement. Right: The retrieved thermal structure for each model. }
    \label{fig:hd209_plot}
\end{figure*}
\subsubsection{Scenario 1}
Here we outline the results for Scenario 1 (WFC3 + Spitzer 3.6 and 4.5 microns). We present the retrieved abundances and reduced $\chi^2$ fits\footnote{We define our reduced $\chi^2$ as $\chi^2$ divided by the number of data points. } ($\chi^2$/n) in Table \ref{tab:scenario1_table}. The cloud free model is the best fit model to the data with a $\chi^2$ of 0.69. Comparing the Bayesian evidence of the cloudy models with the cloud free model shows no support for the inclusion of clouds to describe the data. Therefore, it can be concluded that the data can be described by a cloud free atmosphere with a log(H$_2$O) = -4.39$^{+1.29}_{-0.36}$, this is consistent with \citet{Line2016} who find that H$_2$O is the only molecule robustly detected with an abundance centered around solar value with a 1-sigma range of -3.29 to -4.40.


\subsubsection{Scenario 2}
Here we outline the results for Scenario 2 (WFC3 + Spitzer 3.6, 4.5, 5.6 and 7.8 microns). We present the retrieved abundances and reduced $\chi^2$ fits ($\chi^2$/n) in Table \ref{tab:scenario2_table}. Conversely to Scenario 1, the scattering cloud model provides the best fit to the data with a reduced $\chi2$ of 0.63, however the Bayesian model comparison shows that there is no evidence to support the inclusion of clouds in this data analysis. Hence, we can conclude that the model the best describes the data is one that is cloud free and has a log(H$_2$O) = -4.33$^{+1.23}_{-0.39}$, again consistent with the abundance found in \citet{Line2016}.


We present the best fitting spectra and temperature profiles for each of the retrieved models in Figure \ref{fig:hd209_plot}. It can be seen that all the models are consistent within the error bars. The thermal structure of each model is also consistent with each other.

\subsection{WASP-43b}

\begin{figure*}
    \centering
    \includegraphics[width=\textwidth]{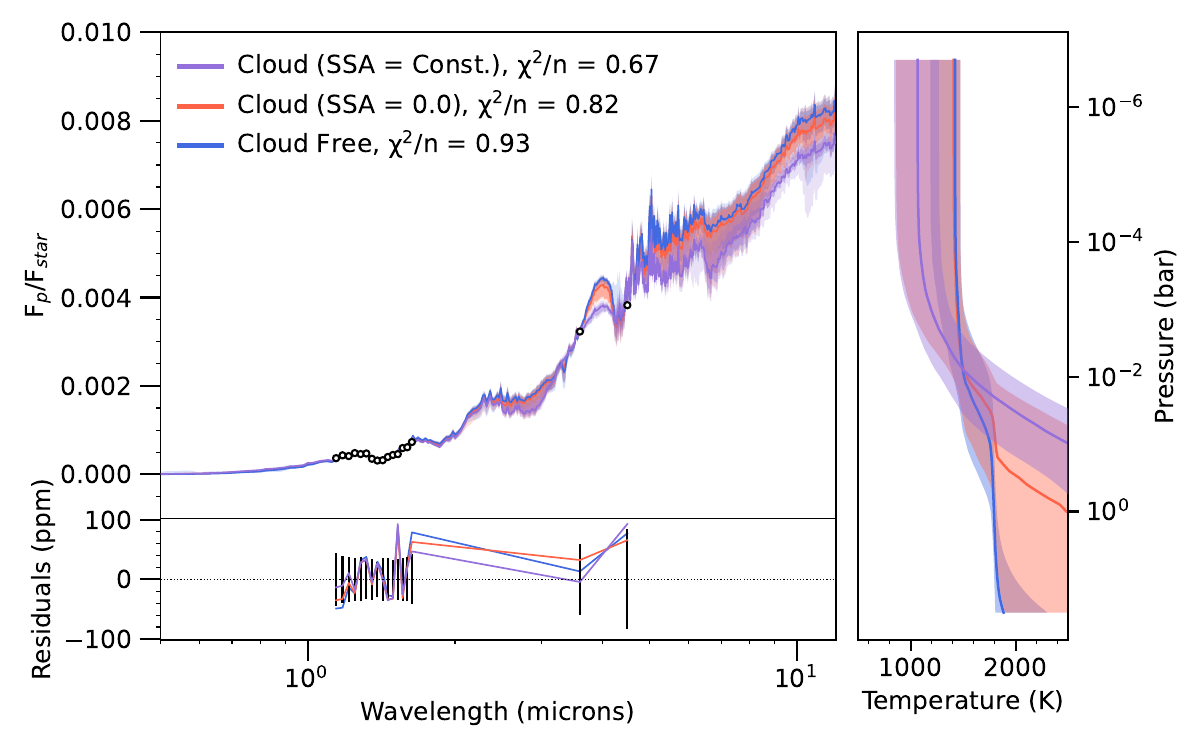}
    \caption{Left: Best fitting models of WASP-43b using the data from \citet{Stevenson2017}. We present the Scat. Cloud, Grey Cloud and Cloud Free models in purple, red and blue respectively. We present the reduced $\chi^2$ for each model inset. The bottom panel shows the residual of the fit, with the black vertical lines representing the error on the measurement. Right: The retrieved thermal structure for each model. }
    \label{fig:stevenson_plot}
\end{figure*}

\begin{figure*}
    \centering
    \includegraphics[width=0.49\textwidth]{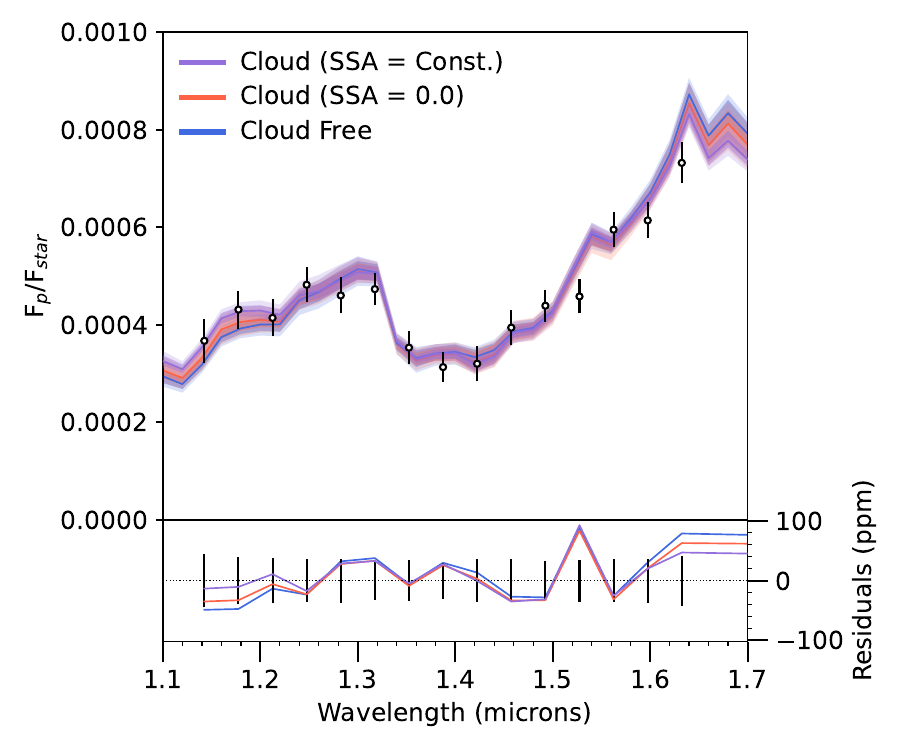}
    \includegraphics[width=0.49\textwidth]{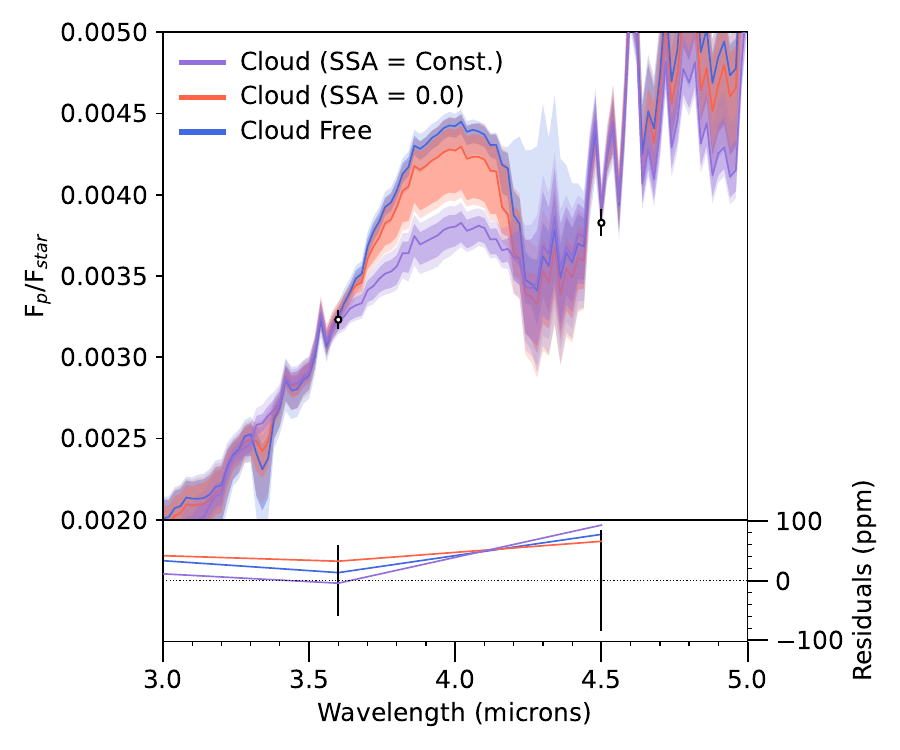}
    \caption{Zoomed plots of the best fitting spectra shown in Fig \ref{fig:stevenson_plot}. Left: The best fitting models to the WFC3 observations. Right: The best fitting models to the Spitzer observations.}
    \label{fig:wfc3_and_spitzer}
\end{figure*}

\subsubsection{Spitzer data from \citet{Blecic2014}}
We present the retrieved chemical abundances for the \textit{Blecic} observations in Table \ref{tab:blecic_table}. The scattering cloud model provides the best fit to the data with a $\chi^2$ of 0.79. However, comparing the Bayesian Evidence of this models with the cloud free model results in a Bayes factor consistent with a non detection. Therefore, the \textit{Blecic} do not support WASP-43b having dayside clouds. We find that a cloud free model with a water abundance of log(H$_2$O) = -3.03$^{+1.03}_{-0.98}$ to be the best model to describe these data. It can be seen from Figure \ref{fig:wasp-43b_data} that the \textit{Blecic} observations have the largest uncertainties, which could result in the model differences being smaller than the uncertainties on the data.


\subsubsection{Spitzer data from \citet{Morello2019}}
We present the retrieved chemical abundances for the \textit{Morello} observations in Table \ref{tab:morello_table}. The scattering cloud model has the best reduced $\chi^2$ with a value of 0.68, however the computed Bayes factor suggests there is no evidence for clouds. Hence we can conclude that these data support a cloud free dayside with log(H$_2$O) = -2.91$^{+0.97}_{-1.06}$ .


\subsubsection{Spitzer data from \citet{May2020}}
We present the retrieved chemical abundances for the \textit{May} observations in Table \ref{tab:may_table}. The scattering cloud model best fit the data, with a reduced $\chi^2$ of 0.67. The scattering cloud model has a Bayes factor of 3.5, which suggests support for clouds on the dayside of the planet. This Bayes factor is converted to 3.13 $\sigma$. The model retrieves a single scattering albedo of 0.66$^{+0.17}_{-0.18}$ and log(H$_2$O) = -2.59$^{+0.68}_{-0.77}$.


\subsubsection{Spitzer data from \citet{Stevenson2017}}
We present the plot of best fitting models to the observations in Figure \ref{fig:stevenson_plot}, with a zoomed in look in Figure \ref{fig:wfc3_and_spitzer}. We present the retrieved chemical abundances for the \textit{Stevenson} simulations in Table \ref{tab:stevenson_table}.  The scattering cloud model best describe the data with a $\chi^2$ of 0.67. Compared to the cloud free model, the scattering cloud model has a Bayes factor of 4.16, which suggests support for clouds on the dayside of the planet. This Bayes factor is converted to 3.36 $\sigma$. The model retrieves a single scattering albedo of 0.66$^{+0.16}_{-0.17}$ and log(H$_2$O) = -2.59$^{+0.68}_{-0.77}$. The similarities of the \textit{May} results with the \textit{Stevenson} results suggests that the 3.6 micron data point is influencing the retrieved results more than the 4.5 micron result.

\subsection{Comparison to Literature}
\begin{figure}
    \centering
    \includegraphics[width=0.49\textwidth]{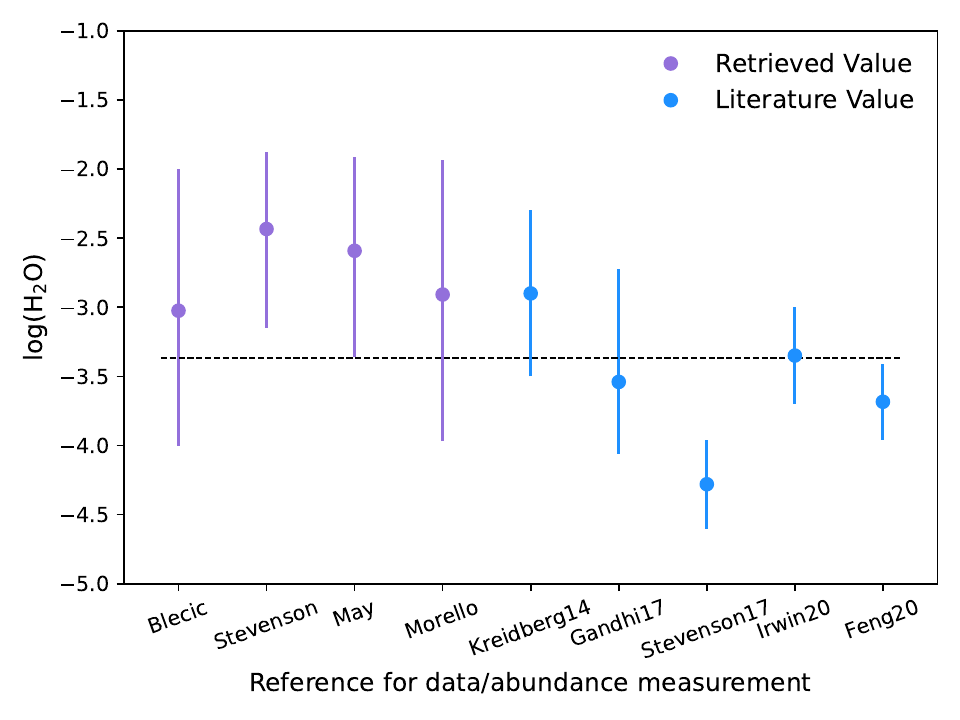}
    \caption{Retrieved H$_2$O abundances from the 4 different data sets used in this study compared to current literature values. The H$_2$O value presented is from the best fit model for each data set. For \textit{Blecic} and \textit{Morello} the best fit model was no offset cloud free and for \textit{May} and \textit{Stevenson} it was no offset scattering cloud. The literature values were obtained from \citet{Kreidberg2014}, \citet{Gandhi2018}, \citet{Irwin2020} and \citet{Feng2020}. The dotted horizontal line represents the solar abundance of H$_2$O at 1400K.}
    \label{fig:h2o_offset}
\end{figure}

The dayside spectra of WASP-43b has been extensively studied and the abundance of H$_2$O has been retrieved by different research groups \citep{Kreidberg2014,Gandhi2018,Stevenson2017,Irwin2020,Feng2020}. In Figure \ref{fig:h2o_offset} we present a comparison of the retrieved H$_2$O abundances from this study to values found in the literature. We find that regardless of the Spitzer data points used, the retrieved abundance of H$_2$O is consistent within 1-$\sigma$ for each data set used and for the majority of the literature values, this is likely driven by the WFC3 data points.

\subsection{The Observable Impact of Dayside Clouds}
\begin{figure*}
    \centering
    \includegraphics[width=\textwidth]{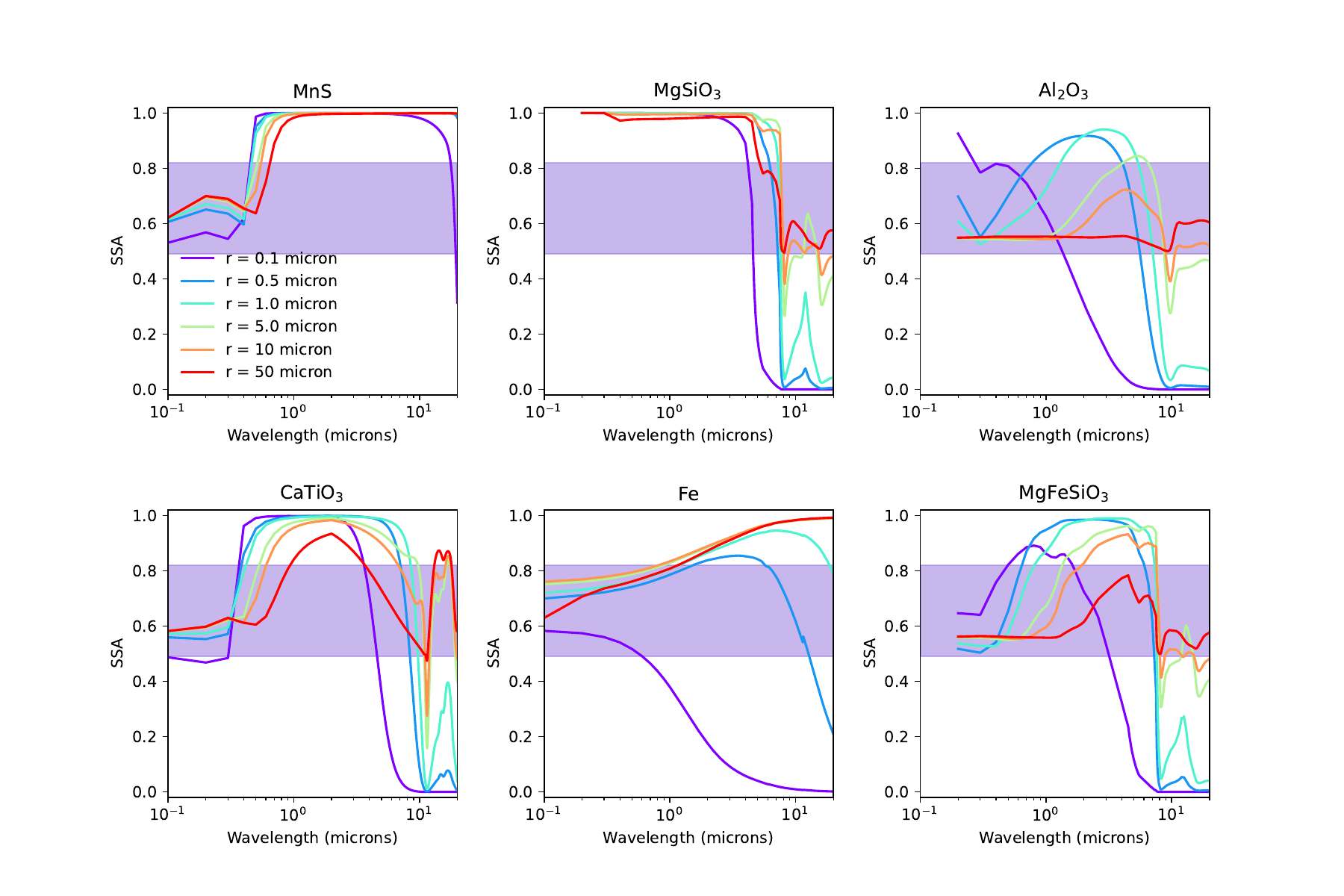}
    \caption{The single scattering albedo spectra for an array of common cloud condensing species with a range of particle sizes. The shaded purple region indicates the best fit SSA from the \citet{Stevenson2017} data.}
    \label{fig:comparessa}
\end{figure*}

The data which provides the most support for dayside clouds are from \cite{Stevenson2017} and the best fit models are shown in Figure \ref{fig:stevenson_plot}. The scattering cloud models are shown in purple and it can be seen that it exhibits lower flux between the Spitzer points compared to the models with no scattering. In Figure \ref{fig:comparessa} we compare the retrieved single scattering albedo to an array of molecules, with the retrieved single scattering albedo value in purple. We can see that we cannot tell what sort of clouds are present. The deviation from the no scattering model would be quantifiable with JWST, hence a definitive detection of dayside clouds would be best made using either NIRSpec/G395H or MIRI/LRS. This is because the impact of the scattering clouds is larger at longer wavelengths. By retrieving the cloud single scattering albedo, one could determine the potential cloud condensing species by comparing to the grid of computed single scattering albedos in Figure \ref{fig:comparessa}. This theory will soon be testable as a phase curve of WASP-43b with NIRSpec/G395H and MIRI/LRS will be observed during Cycle 1.

\section{Conclusions} \label{sec:conclusions}
We performed a reanalysis of the published dayside emission spectra of HD 209458b and WASP-43b. We aimed to explore if these exoplanets contain dayside clouds. We found that there is no evidence for dayside clouds in HD 209548b, which is consistent with previous analysis of this object \citep{Line2016}. We find consistent log(H$_2$O) values as the literature.

The picture becomes more complex when considering WASP-43b as there have been multiple data reductions performed on the Spitzer observations, therefore we analysed each. We found that the detection of scattering clouds was data dependent. Using the Spitzer data from \textit{Stevenson} and \textit{May} produced a Bayes factor of 4.16 and 3.5 respectively, suggesting substantial evidence for clouds according to the Kass and Rafferty scale \citep{kass1995bayes}. Using the formulation from \citet{Trotta2008}, these Bayes factors are converted to 3.36 and 3.13 $\sigma$ respectively. On the other hand, using the Spitzer data from \textit{Blecic} and \textit{Morello} find no evidence for any clouds, with a cloud free model describing the data the best. Despite this, the retrieved log(H$_2$O) was consistent for each best fitting model, hence the abundance of H$_2$O is robust against model assumptions and data reductions.

The inference of clouds was only seen in half of the data sets explored, therefore we can conclude that the data quality is probably not good enough to say anything meaningful about the clouds, if we believe them to be there. With the launch of JWST, this question will be definitively answered.




\section*{Acknowledgements}

Jake Taylor thanks the Canadian Space Agency and the John Fell Fund for the financially supporting this work. We thank the anonymous reviewer for their careful read of this manuscript and for providing comments which greatly improved the science.

\section*{Data Availability}
The models generated in this paper are available on request.



\bibliographystyle{mnras}
\bibliography{bib} 



\appendix
\section{Table of results for each retrieval scenario}

\begin{table*}
    \centering
    \begin{tabular}{cccccccccc}
    & & & & & &\\
    \hline
    Model & logH$_2$O & logCO & logCO$_2$ & logCH$_4$ & logNH$_3$ & SSA & $\chi^2$/n&logE&$\Delta$logE\\
    \hline
    Scat. Cloud     & -3.82$^{+1.66}_{-0.73}$ &-5.25$^{+2.45}_{-4.03}$ &-4.81$^{+2.08}_{-2.54}$ &-8.69$^{+2.08}_{-1.99}$ &-8.45$^{+2.31}_{-2.13}$& 0.52$^{+0.29}_{-0.31}$&0.80 &-16.67 & 0.32\\
    Grey Cloud     & -4.23$^{+1.49}_{-0.43}$ &-5.71$^{+2.70}_{-3.90}$ & -5.72$^{+2.30}_{-2.89}$&-9.07$^{+1.99}_{-1.80}$ &-8.61$^{+2.14}_{-2.01}$&-& 0.71 & -16.74&0.25 \\
    Cloud Free     & -4.39$^{+1.29}_{-0.36}$ &-6.10$^{+3.00}_{-3.52}$ &-7.09$^{+2.90}_{-3.05}$ &-9.09$^{+2.07}_{-1.84}$ &-8.80$^{+2.22}_{-2.10}$&-& 0.69 & -16.99& -\\
    \end{tabular}
    \caption{Table showing the retrieved results from the first HD209458b scenario in which the 5.6 and 7.8 micron data points were not used. We present the model type, retrieved chemical abundances, the single scattering albedo of the scattering cloud, reduced $\chi^2$ of the fit, the log evidence and the Bayes factor, which is the difference between the specific model and the cloud free model.}
    \label{tab:scenario1_table}
\end{table*}


\begin{table*}
    \centering
    \begin{tabular}{cccccccccc}
    & & & & & &\\
    \hline
    Model & logH$_2$O & logCO & logCO$_2$ & logCH$_4$ & logNH$_3$ & SSA & $\chi^2$/n&logE&$\Delta$logE \\
    \hline
    Scat. Cloud     & -4.55$^{+0.43}_{-0.23}$ & -5.62$^{+1.56}_{-3.52}$& -7.64$^{+1.34}_{-2.65}$& -9.33$^{+1.60}_{-1.63}$& -8.82$^{+1.99}_{-2.02}$& 0.48$^{+0.31}_{-0.29}$ &0.63&-18.53&0.34\\
    Grey Cloud     & -4.50$^{+0.71}_{-0.27}$ & -5.78$^{+1.74}_{-3.65}$&-7.39$^{+1.30}_{-2.72}$ & -9.30$^{+1.69}_{-1.76}$& -8.97$^{+2.10}_{-1.92}$& - &0.66& -18.50&0.37\\
    Cloud Free     & -4.33$^{+1.23}_{-0.39}$ & -5.47$^{+1.85}_{-3.98}$& -7.38$^{+1.24}_{-2.89}$& -9.24$^{+2.00}_{-1.81}$& -8.76$^{+2.26}_{-2.17}$& - &0.69& -18.87&-\\
    \end{tabular}
    \caption{Table showing the retrieved results from the second HD209458b scenario in which the 5.6 and 7.8 micron data points were used. We present the model type, retrieved chemical abundances, the single scattering albedo of the scattering cloud, reduced $\chi^2$ of the fit, the log evidence and the Bayes factor, which is the difference between the specific model and the cloud free model.}
    \label{tab:scenario2_table}
\end{table*}

\begin{table*}
    \centering
    \begin{tabular}{ccccccccccc}
    & & & & & & &\\
    \hline
    Model & logH$_2$O & logCO & logCO$_2$ & logCH$_4$ & logNH$_3$ & SSA &$\chi^2$&$\chi^2$/n&logE&$\Delta$logE \\
    \hline
    Scat. Cloud     & -3.07$^{+1.03}_{-0.71}$ &-6.35$^{+3.32}_{-3.51}$ &-4.61$^{+1.22}_{-1.63}$ &-8.12$^{+2.19}_{-2.24}$ &-8.17$^{+2.35}_{-2.26}$ & 0.66$^{+0.25}_{-0.31}$&12.86& 0.76& -21.63&1.23\\
    Grey Cloud     & -2.58$^{+1.04}_{-1.10}$ &-5.56$^{+2.85}_{-3.91}$ &-4.51$^{+1.21}_{-2.98}$ &-7.60$^{+2.38}_{-2.57}$ &-7.49$^{+2.64}_{-2.67}$ & - &16.40&0.96&-22.35& 0.51\\
    Cloud Free     & -3.03$^{+1.03}_{-0.98}$ & -5.57$^{+1.98}_{-3.68}$&-3.96$^{+1.29}_{-0.61}$ &-7.37$^{+2.21}_{-2.88}$ &-7.49$^{+2.73}_{-3.00}$ & - &17.02& 1.00& -22.86&-\\
    & & & & & & &\\
    \hline
    \end{tabular}
    \caption{Table showing the retrieved results from the WASP-43b scenario in which the data from \citet{Blecic2014} was used. We present the model type, retrieved chemical abundances, the single scattering albedo of the scattering cloud, $\chi^2$, reduced $\chi^2$ of the fit, the log evidence and the Bayes factor, which is the difference between the specific model and the cloud free model.}
    \label{tab:blecic_table}
\end{table*}

\begin{table*}
    \centering
    \begin{tabular}{ccccccccccc}
    & & & & & &&\\
    \hline
    Model & logH$_2$O & logCO & logCO$_2$ & logCH$_4$ & logNH$_3$ & SSA&$\chi^2$&$\chi^2$/n&logE&$\Delta$logE \\
    \hline
    Scat. Cloud     &-2.54$^{+0.66}_{-0.98}$  &-4.12$^{+1.83}_{-4.29}$ &-6.01$^{+1.67}_{-3.50}$ &-8.54$^{+2.22}_{-2.07}$ &-8.21$^{+2.50}_{-2.33}$ &0.53$^{+0.28}_{-0.25}$ &11.64&0.68& -21.86&2.95\\
    Grey Cloud     &-2.38$^{+0.53}_{-0.82}$  &-3.78$^{+1.58}_{-4.54}$ &-6.41$^{+2.12}_{-3.57}$ &-7.75$^{+2.85}_{-2.65}$ &-7.80$^{+2.62}_{-2.61}$ &-&12.85& 0.76&-23.21&1.60\\
    Cloud Free     &-2.91$^{+0.97}_{-1.06}$  &-5.43$^{+2.92}_{-4.44}$ &-5.56$^{+1.91}_{-3.98}$ &-5.00$^{+0.97}_{-1.06}$ &-7.48$^{+2.60}_{-2.97}$ &-&15.92& 0.94& -24.81&-\\
    & & & & & &&\\
    \end{tabular}
    \caption{Table showing the retrieved results from the WASP-43b scenario in which the data from \citet{Morello2019} was used. We present the model type, retrieved chemical abundances, the single scattering albedo of the scattering cloud, $\chi^2$, reduced $\chi^2$ of the fit, the log evidence and the Bayes factor, which is the difference between the specific model and the cloud free model.}
    \label{tab:morello_table}
\end{table*}

\begin{table*}
    \centering
    \begin{tabular}{ccccccccccc}
    & & & && &&\\
    \hline
    Model & logH$_2$O & logCO & logCO$_2$ & logCH$_4$ & logNH$_3$ & SSA&$\chi^2$&$\chi^2$/n&logE&$\Delta$logE \\
    \hline
    Scat. Cloud     & -2.59$^{+0.68}_{-0.77}$ &-4.56$^{+1.96}_{-4.26}$ &-6.41$^{+1.68}_{-3.30}$ &-8.43$^{+2.14}_{-2.04}$ &-8.05$^{+2.43}_{-2.46}$ &0.66$^{+0.17}_{-0.18}$&11.33& 0.67&-21.82&3.5\\
    Grey Cloud     & -2.63$^{+0.70}_{-0.80}$ & -3.87$^{+1.52}_{-4.83}$&-5.80$^{+2.45}_{-4.02}$ &-5.08$^{+1.26}_{-2.11}$ &-7.99$^{+2.76}_{-2.55}$ & -&14.32&0.84&-24.36&0.96\\
    Cloud Free     & -2.69$^{+0.92}_{-0.96}$ &-5.50$^{+3.13}_{-4.34}$ &-4.78$^{+1.71}_{-3.63}$ &-4.55$^{+0.95}_{-0.89}$ &-7.91$^{+2.97}_{-2.69}$ &-& 15.76&0.93&-25.32&-\\
    & & & & & &&\\
    \end{tabular}
    \caption{Table showing the retrieved results from the WASP-43b scenario in which the data from \citet{May2020} was used. We present the model type, retrieved chemical abundances, the single scattering albedo of the scattering cloud, $\chi^2$, reduced $\chi^2$ of the fit, the log evidence and the Bayes factor, which is the difference between the specific model and the cloud free model.}
    \label{tab:may_table}
\end{table*}

\begin{table*}
    \centering
    \begin{tabular}{ccccccccccc}
    & & & & & &&\\
    \hline
    Model & logH$_2$O & logCO & logCO$_2$ & logCH$_4$ & logNH$_3$ & SSA&$\chi^2$&$\chi^2$/n&logE&$\Delta$logE \\
    \hline
    Scat. Cloud     & -2.59$^{+0.68}_{-0.77}$ &-4.56$^{+1.96}_{-4.26}$ &-6.41$^{+1.68}_{-3.30}$ &-8.43$^{+2.14}_{-2.04}$ &-8.05$^{+2.43}_{-2.46}$ &0.66$^{+0.16}_{-0.17}$&11.33& 0.67&-21.71&4.16\\
    Grey Cloud     & -2.51$^{+0.73}_{-0.90}$ & -6.70$^{+3.04}_{-3.26}$&-4.90$^{+1.77}_{-0.97}$ &-4.98$^{+1.31}_{-2.62}$ &-7.87$^{+2.72}_{-2.61}$ &-&14.02& 0.82&-24.42&1.48\\
    Cloud Free     & -2.68$^{+0.99}_{-1.00}$ &-6.10$^{+3.31}_{-3.92}$ &-4.62$^{+1.54}_{-2.62}$ &-4.53$^{+1.00}_{-0.92}$ &-7.84$^{+2.87}_{-2.67}$ &-&15.79& 0.93&-25.90&-\\
    & & & & & &&\\
    \end{tabular}
    \caption{Table showing the retrieved results from the WASP-43b scenario in which the data from \citet{Stevenson2017} was used. We present the model type, retrieved chemical abundances, the single scattering albedo of the scattering cloud, $\chi^2$, reduced $\chi^2$ of the fit, the log evidence and the Bayes factor, which is the difference between the specific model and the cloud free model.}
    \label{tab:stevenson_table}
\end{table*}
\bsp	
\label{lastpage}
\end{document}